Electronic Access to Information and the Privacy Paradox: Rethinking "Practical Obscurity"

A Paper Presented to TPRC: The 29th Annual Conference on Information, Communication and Internet Policy

Oct. 27-29, 2001

By

Charles N. Davis, Ph.D.

Director, Freedom of Information Center

Missouri School of Journalism

University of Missouri

76 Gannett

Columbia, MO 65211

(573) 882-5736

Electronic Access to Information and the Privacy Paradox: Rethinking "Practical Obscurity"

I. Introduction

The decision of the United States Supreme Court in *Department of Justice v. Reporters Committee for Freedom of the Press* dramatically altered the treatment of federal Freedom of Information Act (FOIA)[1] requests implicating privacy interests. Certainly the Court's decision had a dramatic impact on the balancing of privacy interests and the public interest in governmental transparency. The Court's opinion in *Reporters Committee* is most often noted for its conclusion that the FOIA's core purpose was to shed light on governmental activities and operations and for its assertion that disclosure of personal information seldom accomplishes that goal.

Lost in the debate over *Reporters Committee*, however, is the Court's creation of the concept of "practical obscurity," the idea that records once stored in separate repositories scattered throughout the halls of government are transformed by computerization into new record forms that implicate privacy interests in new ways. Indeed, the ease and speed with which the personal computer can synthesize disparate records, and the universal access to the records created by the Internet, create novel issues for freedom of information law. In its initial look at the issue, the *Reporters Committee* Court opted for a bright-line test: information in federal records that is "about the government" is public, and information about individuals is not.

---

[1] See 5 U.S.C. § 552 (1994).

As the cases following *Reporters Committee* have shown, records rarely fall into such neat categories. In fact, information about individuals can, and often does, shed light on governmental operations and activities. Since *Reporters Committee*, lower courts facing the issue of access versus privacy have almost universally found that records documenting governmental involvement in ways that involve personal information about those involved in government programs, or even federal investigations of alleged wrongdoing, should be protected by the privacy Exemption 6 or by Exemption 7(c) of the FOIA, which applies privacy law to law enforcement records. The result is a body of law that treats information as if it falls into one category or the other, when nothing could be further from reality.

At the heart of this doctrine is the notion of practical obscurity, the Court's belief that public records should be public so long as they are separated by geographic distance and antiquated paper-based systems. Practical obscurity also gives rise to the assumption that computerized compilations of once-obscure records transforms public records into private data thanks merely to the ease with which such records can now be accessed.

This paper will demonstrate how the Court's decision in *Reporters Committee* -- specifically its formulation of the "central purpose" of the FOIA[2] and its embrace of practical obscurity -- holds far-reaching implications for federal access law. Examination of *Reporters Committee* and of more than 60 lower federal court opinions relying in whole or in part upon the Court's "central purpose" formulation in *Reporters Committee* show that the Court has sharply limited the ambit of information that can be released under the Freedom of Information Act.[3] Finally, this paper will offer an alternative to the *Reporters Committee* doctrine created through a line of cases addressing similar issues in the context of the Privacy Act, a close legislative cousin of the FOIA, that offers a compromise ofسorts between privacy and access.

II. Department of Justice v. *Reporters Committee for Freedom of the Press*

*Reporters Committee* presented a question crucial to the emergence of electronic access: Could an agency invoke privacy concerns to deny access to public records compiled electronically, for no other reason than that they were compiled electronically? Does the mere compilation of data give rise to otherwise non-existent privacy concerns? In addressing these questions, the Court issued its first opinion on the relationship between access and privacy, indicating a clear preference for a narrower application of the FOIA. The case arose after a reporter filed a FOIA request[4] asking for the FBI's "rap sheet" on Charles Medico, a businessman identified by the Pennsylvania Crime Commission as an owner of Medico Industries, a legitimate business dominated by organized-crime figures.[5] The reporter was investigating Medico because Medico's company received defense contracts allegedly in exchange for political contributions to U.S. Rep. Daniel J.

---

[2] See Department of Justice v. Reporters Committee for Freedom of the Press, 489 U.S. 749, at 774.

[3] See S. REP. NO. 104-272, at 26-27 (Statement by Sen. Leahy)(1996); Reporters Comm. for Freedom of the Press, Report on Responses and Non-Response of the Executive and Judicial Branches to Congress' Finding that the FOI Act Serves 'Any Purpose' 2-3 (July 2, 1998).

[4] See 489 U.S. at 757.

[5] See id.

Flood.[6] Flood, who eventually left office in disgrace, already was under investigation for corruption.[7]

      The FBI released information on three of Charles Medico's brothers, all deceased, but the agency refused to release Charles Medico's rap sheet on privacy grounds because he was still alive.[8] The reporter sued to gain access to the records, but the U.S. District Court for the District of Columbia granted the FBI's motion for summary judgment to dismiss the suit. The court held that the information was protected under the privacy provision of the FOIA's law enforcement exemption,[9] and that disclosure of the records would be an unwarranted invasion of Charles Medico's privacy.[10]

On appeal, however, the U.S. Court of Appeals for the D.C. Circuit ruled in favor of the CBS journalist and Reporters Committee. The appeals court reasoned that the government can not claim a privacy interest in an FBI compilation of law enforcement agency records when those same records would be available as public records from the individual agencies themselves.[11]

---

[6]See id.

[7]See Laura Kiernan, Flood Is Placed on Year's Probation, THE WASH. POST, Feb. 27, 1980, at A8.

[8]See 489 U.S. at 757.

[9]See 5 U.S.C. § 552 (b)7(C)(1994). The exemption states that the FOIA does not apply to matters that are "(7) records or information compiled for law enforcement purposes, but only to the extent that the production of such law enforcement records or information (C) could reasonably be expected to constitute an unwarranted invasion of personal privacy. Exemption 7(C) is one of two privacy exceptions to the FOIA. The other exception, Exemption 6, pertains to "personnel and medical files and similar files the disclosure of which would constitute a clearly unwarranted invasion of personal privacy." See id. § 552 (b)6.

[10]See 489 U.S. at 757-59.

[11]See Reporters Comm. for Freedom of the Press v. U.S. Dep't of Justice, 816 F.2d 730, 740 (1987).

The Department of Justice appealed to the Supreme Court, which balanced the individual's right of privacy against the public interest in disclosure and reversed the appellate court ruling, thus allowing the FBI to withhold the information.[12] Writing for the Court, Justice John Paul Stevens said the FOIA's "central purpose is to ensure that the government's activities be opened to the sharp eye of public scrutiny, not that information about private citizens that happens to be in the warehouse of the government be so disclosed."[13] The Court reasoned that because a computerized compilation of an individual's rap sheet does not directly shed light on governmental performance, it falls "outside the ambit of the public interest that the FOIA was enacted to serve."[14] The requested FBI records would "tell us nothing directly about the character of the Congressman's behavior," Justice Stevens wrote.[15] "Nor would it tell us anything about the conduct of the Department of Defense in awarding one or more contracts to the Medico Company."[16]

Certainly one could argue that Medico's rap sheet would provide details to embolden a news story, but the Court reasoned that this is not the kind of public interest for which Congress enacted the FOIA. So, while doubtless there exist some public interest in anyone's criminal history, especially if that history details the subject's relationship with governmental officials or agencies, the FOIA's central purpose -- scrutiny of government -- forestalls such inquiry.

---

[12]See 489 U.S. at 772-73 (citing 425 U.S. at 372.)

[13]See 489 U.S. at 774.

[14]See id. at 775.

[15]See id. at 774.

[16]See id.

Of particular interest to this paper is the Court's treatment of the technology involved in compilations. According to the Court, a citizen possesses a protected privacy interest in the criminal history information because "plainly there is a vast difference between the public records that might be found after a diligent search of courthouse files, county archives, and local police stations throughout the country and a computerized summary located in a single clearinghouse of information."[17] Thus, individuals maintain a privacy interest in the "practical obscurity" of records.[18]

Taken literally, this means that public documents that are difficult or time-consuming to locate essentially become private records. And when those hard-to-find documents are assembled electronically, they implicate privacy interests in new ways. Writing shortly after the Court's decision, the noted FOIA expert Harry A. Hammitt posed the following scenario:

Say, for example, that a news organization sent reporters to courthouses and police stations around the country to retrieve public criminal-history records -- arrests, convictions, and sentences.

The news organization could then computerize the information so reporters could enter a name and call up a profile of that individual's criminal history. Since all the information is admittedly public, such a private database would not invade anyone's privacy, at least not legally.

---

[17] Id. at 764.

[18] Id.

But when government does the same thing with its computers, according to the Court's ruling, an overarching concern for privacy suddenly sweeps away the public nature of the information and dramatically restricts the scope of the FOIA.[19]

The restructuring of the FOIA in Reporters Committee through the lens of the "core purpose" test, coupled with the Court's creation of "practical obscurity," clearly demonstrates that the burden of proof in privacy-related cases now rests with the requester rather than with the government. Instead of a presumption of openness, there now exists a requirement that the requester show that the information sought will reveal -- directly -- something about governmental operations.

Where governmental operations intersect with the lives of citizens -- and they do, daily, at the federal, state and local level -- should individuals reasonably expect that those dealings will always be private? If so, then a prime purpose of the FOIA, the exposure of fraud, waste and abuse by government, has been tossed aside in favor a reflexive privacy right that seemingly has no end in sight.

Predictably enough, lower courts have faithfully followed the Reporters Committee doctrine, holding in a steadily growing line of cases that the "central purpose" test limits the disclosure of a wide variety of documents, most electronically compiled, that identify individuals.[20] Judges used the "central purpose" analysis in FOIA litigation involving the FOIA privacy exemptions -- Exemptions 6 and 7(C)[21] -- because invasion of privacy was the central issue in Reporters Committee. However, several courts have used the Reporters Committee test[22] in FOIA cases involving issues that go beyond the Act's privacy exemptions.[23]

---

[19] Harry A. Hammitt, "High Court Alchemy," The Quill, Vol. 7, No. 9, 28, 29 (Oct. 1989).

[20] See The Reporters Committee for Freedom of the Press, Report on Responses and Non-Response of the Executive and Judicial Branches to Congress' Finding That the FOI Act Serves "Any Purpose," prepared by request of the Chairman, Subcommittee on Government Management, Information and Technology of the House Committee on Government Reform and Oversight, July 2, 1998. See, e.g., Sheet Metal Workers International Ass'n, Local 19 v. U.S. Dep't of Veteran Affairs, 135 F.3d 891 (3d Cir. 1998); Kimberlin v. U.S. Dep't of Justice, 139 F.3d 944 (D.C. Cir. 1998); McQueen v. United States, 179F.R.D.522 (S.D. Tex. 1998); Ligorner v. Reno, 2 F.Supp.2d 400 (S.D.N.Y. 1998); Lurie v. U.S. Dep't of the Army, 970 F. Supp. 19 (D.D.C. 1997); Center to Prevent Handgun Violence v. U.S. Dep't of the Treasury, 981 F. Supp. 20 (D.D.C. 1997); Sheet Metal Workers International Ass'n, Local 9 v. U.S. Air Force, 63 F.3d 994 (10th Cir.

---

1995); Manna v. U.S. Dep't of Justice, 51 F.3d 1158 (3d Cir. 1994), cert. denied, 116 S. Ct. 477 (1995); Exner v. U.S. Dep't of Justice F. Supp. 240 (D.D.C. 1995); Jones v. FBI, 41 F.3d 238 (6th Cir. 1994); U.S. Dep't of the Navy v. Federal Labor Relations Authority, 975 F.2d 348 (7th Cir. 1992); Hunt v. FBI, 972 F.2d 286 (9th Cir. 1992); and Hale v. U.S. Dep't of Justice, 973 F.2d 894 (10th Cir. 1992).

[21]See 5 U.S.C. §§ 552(b)6 & (b)7(C)(1994). The statutory language of privacy Exemptions 6 and 7(C) reflect two key differences. First, Exemption 6 calls for "a clearly unwarranted invasion" of privacy (italics added). See id. Exemption 7 requires a less strict standard, asking an agency to show only "an unwarranted invasion of privacy." See id. § 552 (b)7(C). Second, Exemption 6 applies to information that, if disclosed, "would constitute" an invasion of privacy (italics added). See id. § 552 (b)6. Exemption 7, on the other hand, applies to information the disclosure of which "could reasonably be expected to constitute" an invasion of privacy (italics added). See id. § 552 (b)7(C).

 The legislative history shows that the difference in language was intentional. Exemption 7(C), as originally proposed by Sen. Gary Hart, also required a "clearly" unwarranted invasion of personal privacy. See 120 CONG. REC. 17033 (1974). However, the word "clearly" was dropped by the Conference Committee as a concession in negotiations with President Ford to get the Act approved. See CONF. REP. NO. 93-1380, at 11 (1974). By dropping "clearly," the Exemption lessened the agency's burden to meet the test. See JAMES T. O'REILLY, FEDERAL INFORMATION DISCLOSURE: PROCEDURES, FORMS AND THE LAW § 17.09, at 13-44 (1994). Legislators also agreed to the difference in language between "would" in Exemption 6, and "could reasonably be expected" in Exemption 7(C) in order enact the legislation. Courts have consequently concluded that Exemption 7(C) allows law enforcement officers more latitude to withhold records to protect privacy than is permitted under the stricter standard of Exemption 6. See U.S. Dep't of Justice v. Reporters Comm. for Freedom of the Press, 489 U.S. 749, 755-756 (1989). In addition, Exemption 7(C) means the public interest in disclosure carries less weight. See id.

 When judges make a determination in a privacy-interests case under Exemption 7(C), the courts use a two-step test. See 5 U.S.C. § 552(b)7(C). First, the documents must have been compiled for law enforcement reasons because this Exemption pertains only to investigative records. Second, the government must prove that the disclosure could "reasonably be expected to constitute an unwarranted invasion of privacy." See id. Similarly, the courts use a similar test in deciding an Exemption 6 privacy-interests case. The courts first must determine if the records falls within the definition of "personnel," "medical" or "similar" files. See id. § 552(b)6. Second, the courts must balance the invasion of the individual's personal privacy against the public benefit that would result from disclosure. To withhold information, the government must show that the disclosure "would constitute a clearly unwarranted invasion of privacy." See id. See also U.S. Dep't of the Air Force v. Rose, 425 U.S. 352, 382 (1972).

[22]See 489 U.S. at 774-75.

[23]See Christopher P. Beall, The Exaltation of Privacy Doctrines Over Public Information Law, 45 DUKE L.J. 1249, 1273 (1996). Beall cited three cases in which courts have broadened the applicability of the central purpose doctrine. See id. at 1273-80. See Sweetland v. Walters, 60 F.3d 852 (D.C. Cir. 1995)(per curiam)(holding that the Executive Residence staff of the White House is not an "agency" under the FOIA); Baizer v. U.S. Dep't of the Air Force, 887 F. Supp. 225 (N.D. Cal. 1995)(holding that an electronic copy of the Air Force's computerized database of Supreme Court opinions should not be considered an "agency record" under the FOIA); Vazquez-Gonzalez v. Shalala, Civ. No. 94-2100 (SEC)(D.P.R. Feb. 13, 1995)(dismissing a suit brought by a physician, who sought information about Medicare billing practices,

In all of these cases, courts after Reporters Committee have closely scrutinized public interest assertions. The federal courts narrowly define the "central purpose" of the FOIA as records that will shed light on agency performance directly. The analysis does not extend to the question of whether the records will facilitate reportage, and instead limits the public interest inquiry to the contents of the records themselves. Indeed, where documents have been sought largely for the personal information they contained, the courts have been protective of privacy interests unless the public interest is overriding. In fact, one federal court has said that the public interest argument is insubstantial unless the requester "puts forward compelling evidence that the agency denying the FOIA request is engaged in illegal activity and shows that the information sought is necessary to confirm or refute that evidence."[24]

The concept of "practical obscurity" as an interest mitigating toward privacy is seldom discussed in these post-Reporters Committee cases, but the notion certainly thrives within the "central purpose" doctrine. The cases present an overall picture of a federal judiciary actively reigning in citizen access to the information collected and compiled by government. The fact that much of that information is compiled, stored and disseminated through computer networks leads to the logical conclusion that the more government documents created electronically, the greater the "practical obscurity" of the

---

because the requested information concerned the plaintiff's own commercial interests). See Beall, at 1273-80.

[24] See Computer Professionals for Social Responsibility v. Secret Service, 72 F.3d 897, 904-05 (D.C.Cir. 1996). See also Accuracy in Media, Inc. v. National Park Service, 194 F.3d 120, 124 (D.C. Cir. 1999), cert. denied, 68 U.S.L.W. 3711 (May 15, 2000) (inconsistencies in crime scene reports are "hardly so shocking as to suggest illegality or deliberate government falsification."); SafeCard Services Inc. v. SEC, 926 F.2d 1197 (D.D.C. 1991) ("Indeed, unless there is compelling evidence that the agency denying the FOIA request is engaged in illegal activity…there is no reason to believe that the incremental public interest in such information would ever be significant.")

data. Despite the revolutionary ability of the computer, and of the Internet, to make information more readily available to the citizenry, "practical obscurity" stands ready to limit its vast potential to democratize information.

At the heart of "practical obscurity" is the Court's new categorical approach in cases involving privacy claims. To the Reporters Committee Court, information is either about individuals or about government; when the two categories blend, the result, in the Court's view, should almost always be non-disclosure. Neither life, nor data, is ever quite so simple. A more reasoned approach does exist, however, in a line of cases interpreting the Privacy Act, a statute closely related to the FOIA.

The Privacy Act Approach: When is a record "about" an individual?

Enacted in the wake of the Watergate scandal of 1974, the Privacy Act is an important yet little-known statute.[25] It represents a measured response by Congress to the public's concern that the federal government was gathering vast amounts of data about individuals and that the data might be in ways that threatened personal privacy.[26] The Privacy Act was designed as a protector of personal privacy against the potential government misuse of information.[27] The Act specified that its purpose was "to provide certain safeguards against an invasion of personal privacy," through the fulfillment of six guiding provisions.[28] As a statute aimed at protecting the right of the people to maintain some degree of control over information the government collects about them, it can be viewed as the legislative counter to the FOIA.

---

[25] For commentary on the Privacy Act, see, e.g., Todd Robert Coles, Comment, "Does the Privacy Act of 1974 Protect Your Right to Privacy," 40 Am. U. L. Rev. 957 (1991);

[26] Note, "The Privacy Act of 1974: An Overview and Critique," 1976 Wash. U. L.Q. 667, 669-70.

[27] Cornish F. Hitchcock, "Overview of the Privacy Act," in Justin D. Franklin and Robert E. Bouchard, Guidebook to the Federal Freedom of Information and Privacy Acts, § 2.02, 2-18.3 (Supp. 2001) (hereinafter cited as *Guidebook*).

The Privacy Act of 1974 attempted to strike a delicate balance between the government's need to gather and to use personal information and the individual's competing interest in maintaining control over such personal information. Broadly stated, the Act requires every federal agency maintaining a record on an individual within a system of records to: (1) permit the individual to control the use and dissemination of information contained in the record; (2) permit the individual to review, to correct, or to amend information contained in the record; (3) regulate and restrict the collection, maintenance, use, and dissemination of information in the record; and (4) be subject to civil suit for specified violations of the Privacy Act.[29] Collectively, these safeguards are designed to protect individual privacy, while preserving the government's ability to gather and to use personal information.

A key question under the Privacy Act, then, is what records are subject to the various provisions of the Act. The statute defines a "record" as follows:

> …any item, collection, or grouping of information about an individual that is maintained by an agency, including, but not limited to his education, financial transactions, medical history, and criminal or employment history, and that contains his name, or the identifying number, symbol, or other identifying particular assigned to the individual, such as a finger or voice print or a photograph."[30]

---

[28] The provisions are as follows: 1. Individuals would be permitted to determine what records pertaining to themselves were collected, maintained, used or disseminated by federal agencies. 2. Individuals would be able to prevent records about themselves that an agency had obtained for one purpose from being used for another purpose without the person's prior consent. 3. Individuals could not only gain access to records an agency has on them, but could also have the information corrected or amended. 4. Agencies were required to make sure that information about individuals was current and accurate. 5. Agencies were exempted from the requirements of the Act "only in those cases where there is an important public policy need for such exemption" as specified in the Act. 6. Agencies would be subject to a lawsuit for any damages that occur as a result of willful or intentional acts which violate any person's rights under the Act. See Pub. L. No. 93-579, § 2(b), 88 Stat. 1897 (1974).

[29] 5 U.S.C. § 552a(b) (1988).

Such a definition obviously has a broad sweep, applying not only to records containing a person's name, but also to any other document, regardless of physical form, that could be used to identify someone. In fact, the legislative language accompanying the compromise version of the Act upon its passage specified that the term "record" was defined so as "to assure that the intent that a record can include as little as one descriptive item about an individual."[31]

As further evidence of the breadth of the term "record," the Office of Management and Budget, the agency with executive oversight of the Privacy Act, states in its *Privacy Act Guidelines* that the term means "any" item of information about an individual.[32] The OMB Guidelines point out that prohibitions on the on the disclosure of a record may apply not only to the entire record in its compiled state, but also to any item within that record set, provided that the record set contains an individual identifier.[33]

Consistent with the OMB Guidelines, the Court of Appeals for the Third Circuit held that the term "record" "encompasses any information about an individual that is linked to that individual through an identifying particular."[34] Such a construction resembles the United States Supreme Court's reasoning in *Reporters Committee*: if a record in any way identifies an individual, privacy rights are implicated.

---

[30] 5 U.S.C. § 552a(a)(4) (1988).

[31] 120 Cong. Rec. 40, 408 (1974).

[32] 40 Fed. Reg. 28,948, 28,951 (1975).

[33] 40 Fed. Reg. 28,951 (1975).

[34] Quinn v. Stone, 978 F.2d 126, 133 (3d Cir. 1992).

Perhaps sensing the limitless scope of a Privacy Act guided by such logic, the Court of Appeals for the District of Columbia in 1994 limited its reach by adopting a narrow construction of the term "record." The court's more detailed textual analysis of the Act's definition of "record" in *Tobey v. National Labor Relations Board*[35] suggests that a record may not be "about" someone for purposes of the Act merely because the document contains the person's name or other identifier.

*Tobey* involved a computer system used by the National Labor Relations Board for tracking the progress of its cases. One of the fields in the system was for the initials of the field examiner assigned to each case.[36] Apparently, the board has traditionally used the system to prepare its annual report to Congress.

The plaintiff in *Tobey* worked for the NLRB as a field examiner. In connection with a grievance arbitration, an NLRB official conducted a search of the system using his initials to pull up his caseload, days worked and other information.[37] The plaintiff brought suit in district court against the NLRB under the Privacy Act, alleging that the NLRB maintained and used a system of records to retrieve personal information about him and disclosed that information to others without his consent.[38] The district court found that the NLRB had not met the notice requirements of the Privacy Act but concluded that the data retrieved were not "records" within the meaning of the Act.[39]

---

[35] 40 F.3d 469 (D.C. Cir. 1994).

[36] Id. at 470.

[37] Id. at 471.

[38] Id.

[39] Id., citing Tobey v. NLRB, 807 F. Supp. 798, 800-01 (D.D.C. 1992).

On appeal, Tobey argued that the district court improperly defined "record," and argued that the information disseminated by the NLRB was "about" him because it contained identifiers linking him to the information.[40]

Rather than summarily conclude that any and all information that includes an individual identifier automatically meets the definition of "record" under the Act, the court instead turned to statutory text.[41] First, the records must be "about an individual"; second, they must contain an identifying number, symbol or other particular.[42] The court reasoned that the second requirement leads to an obvious conclusion about the first: the mere inclusion of an identifier does not mean that the information necessarily is "about" the individual. If it did, there would be no need for the first prong of the statute. Thus, the court concluded that in order to meet the statutory definition of "record," the information must include some identifying information that is "about" the individual.[43] This conclusion led the court toward a more substantive evaluation of the nature of the information.

To determine whether the information in the NLRB CHIPS system was "about" Tobey, the court examined the use of the data in question. Because the database was used to maintain status reports on NLRB cases rather than on individual employees, the court concluded that the database held no information about individuals sufficient to trigger Privacy Act protections.[44] Rather, they contained information such as case name, allegations made, the number of private-sector employees involved and the dates of hearings, settlements and dismissals.[45]

---

[40] Id.

[41] Id.

[42] Id. at 470-71. For the statutory language, see 5 U.S.C. § 552a (a) (5) (1998).

[43] Id. at 471.

[44] Id. at 471.

Admittedly, the system also includes the number and initials of the field examiner assigned to the case, but the court reasoned that "this no more means the information is 'about' the individual than it means the information is 'about' the date on which the case is settled.[46] Thus, the court construed the Privacy Act protects information that is "about" an individual, not that which simply "identifies" an individual. The latter category is a broad one, encompassing virtually all data relating to an individual; the former, however, includes only information that "actually describes" the individual.[47] In other words, the mere possibility that NLRB officials could use data from CHIPS in combination with other information to draw inferences about his job performance did not transform the data into "records" protected by the Privacy Act.[48]

*Tobey* has proven influential in shaping this area of Privacy Act law, but the federal courts still are mostly divided over the textual approach in determining when information becomes a "record" for purposes of the Act.[49] For example, in Bechhoefer v. Department of Justice,[50] a citizen wrote to the Drug Enforcement Administration suggesting that individuals connected to a local sheriff's office were involved in narcotics trafficking.

---

[45] Id. 472.

[46] Id.

[47] Id.

[48] Id. at 473.

[49] *See, e.g.*, Fisher v. National Insts. Of Health, 934 F. Supp. 464 (D.D.C. 1996) (legend appended to NIH database were about the articles in the database, and not about the authors); but see also Becchoefer v. Department of Justice, 209 F.3d 57 (2d Cir. 2000), revsg Bechhoefer v. Department of Justice, 934 F. Supp. 535 (W.D.N.Y. 1996) (letter concerning alleged drug dealers was about the author since it contained name, address, telephone number, etc.).

[50] 934 F. Supp. 535 (W.D.N.Y. 1996).

A DEA agent sent the letter to the sheriff, and the author was charged with making false statements (a charge that was later dropped).[51] The author filed suit under the Privacy Act to sanction the release of the letter, but the trial court, relying on *Tobey*, concluded that the letter was not "about" the author, but about the suspected drug dealers.[52]

The Second Circuit reversed the Western District of New York in an opinion critical of the D.C. Circuit's opinion in *Tobey*.[53] The Bechhoefer panel identified three approaches to identifying a "record" under the Privacy Act:[54] (a) *Tobey*'s analysis that the information must actually describe the individual in some way;[55] (b) the more exapnsive view in the Ninth and Eleventh circuits that the information must "reflect some quality or characteristic of the individual involved";[56] and the Third Circuit's definition in a 1992 case that a record encompasses "any information about an individual that is linked to the individual through an identifying particular."[57]

---

[51] Id. at 536-37.

[52] Id. at 538.

[53] For commentary on this line of cases, see Guidebook, § 2.04[3], 2-36, 2-36.1, 2-36.2 (Release #29, Dec. 2000).

[54] Id.

[55] 40 F.3d at 772.

[56] Boyd v. Secretary of the Navy, 709 F.2d 684, 686 (11th cir. 1983); Unt v. Aerospace Corp., 765 F.2d 1440, 1449 (9th Cir. 1985).

[57] Quinn v. Stone, 978 F.2d 126 (3d Cir. 1992).

The *Bechhoefer* panel followed the third approach as most consistent with the statutory definition of "record," concluding that the Act contains no language suggesting that the information must reflect a "quality or characteristic" of the individual.[58] Since the letter at issue contained the plaintiff's name, address and other identifying information, it was deemed a "record" about him.[59]

The conflict in approach between the federal circuit courts in defining "records" under the Privacy Act merits further scrutiny, and could lead the United States Supreme Court to settle the issue in a future case. Indeed, it is unclear whether future cases will be decided under one standard rather than another. The information in *Tobey*, for example -- case-tracking documents -- likely would not meet even the narrower standard of *Bechhoefer* because they did not deal with any personal information about the employee, unlike the letter in *Bechhoefer*, which did.

It is clear, however, that courts construing the Privacy Act have adopted a much more flexible approach in sifting through privacy claims than have courts construing the Freedom of Information Act. While the Privacy Act cases focus on a critical examination of the nature and uses of data, the influence of the *Reporters Committee* Court on the FOIA has created a reactive judicial philosophy that favors closure over disclosure.

The Post-*Reporters Committee* Landscape: *McNamera v. Department of Justice*

---

[58] 209 F.3d 57, at 61.

[59] Id. at 62.

A close look at a typical post-*Reporters Committee* decision illustrates the many problems of FOIA requests implicating privacy interests and offers a framework for applying the textual analysis common to many Privacy Act determinations of when a record is "about" an individual. In *McNamera v. Department of Justice*,[60] a federal district court denied a journalist access to records concerning the investigation and prosecution of a West Texas sheriff in a notorious case, ignoring strong public-interest arguments by relying on the *Reporters Committee* "central purpose" rationale.

In *McNamera*, a journalist sought access to records concerning a rather spectacular story: a sheriff convicted for helping a drug-runner smuggle 2,421 pounds of cocaine -- with a street value of $1.1 billion -- into the United States.[61] Two years after his conviction, the journalist McNamera requested from the Department of Justice records concerning the case. McNamera's request stated that he was especially interested in information that disclosed "agency procedures and the working[s] of government."[62]

Despite the widespread public attention in the case and its rather obvious link to governmental investigatory techniques, the court found that McNamera's request had not raised legitimate questions about government action, stating that in order to trigger the sort of public interest that would outweigh privacy concerns, the request must "put forward compelling evidence" that the agency involved is engaged in illegal activity.[63] To the court, the request for law enforcement records does little, if anything, to advance knowledge of the workings of government, unless the requester could present a *prima facie* case of illegality.[64]

---

[60] 974 F.Supp. 946 (W.Dis. Tex. 1997).

[61] Id. at 949.

[62] Id.

[63] Id. at 960.

McNamera argued that disclosure of the records would shed light on the turf wars of the various federal agencies in West Texas. He noted that the drug-runner was registered as an informant by one agency while he was being investigated by another agency involved in the sting.[65] Significant questions existed concerning the proper role of law enforcement, its spending priorities and its proclivity for high-profile busts.[66]

The court brushed aside these arguments, describing them as a "fishing net" and showing far more concern for the potential "stigma" attached to the sheriff (a federal convict). It concluded that no records should be released. The court's solicitude for criminals is not unusual; *Reporters Committee* gives great deference to the privacy rights of convicted felons. The *McNamera* court's narrow view of the public interest in an investigation featuring a billion dollars' worth of cocaine and a corrupt sheriff also is typical of post-*Reporters Committee* privacy discussions, in which courts attempt to divorce personal information from government information and treat the categories as if they never overlap. This categorical approach defies logic, as personal information rarely can be neatly separated from information about government operations.

Such an analysis appears counter to the analysis under the Privacy Act concerning whether a record is "about" an individual. Under the various approaches to this question in the Privacy Act cases, it is clear that personal information -- particularly in the context

---

[64] Id.

[65] Id.

[66] "Court Finds No Public Interest in Drug Investigation," *Access Reports*, August 20, 1997, p.1

of criminal investigations, which depend for their very existence upon personal information -- can be "about" governmental operations as well as "about" individuals.

Where the lines between personal and governmental information blur, there must be some judicial consideration given to the purpose of the FOIA, as defined by the *Reporters Committee* court: "that the government's activities be opened to the sharp eye of public scrutiny, not that information about private citizens that happens to be in the warehouse of the government be so disclosed."[67] The information at issue in *McNamera* clearly does not *happen* to be in the warehouse of the government. It exists because it is *about* the government's investigation and subsequent conviction of a renegade sheriff for a multimillion-dollar drug business.

Conclusion

The *Reporters Committee* "central purpose" test, as interpreted by the lower federal courts, creates the [perhaps] unintended consequence of forestalling any judicial analysis into whether the information sought is intended to further knowledge about governmental operations. Rather than require courts to engage in the hard work of determining whether information sought through FOIA is predominantly "about" government or predominantly "about" individuals, the *Reporters Committee* Court offers an either/or proposition. The result is a test that robs the FOIA of much of its impact by reflecting the Court's dubious assertion that where governments and individuals collide, privacy interests must always triumph. In fact, the opposite may be true, for where governments involve themselves directly in the lives of citizens, the societal interest in oversight might be at its greatest.

---

[67]See 489 U.S. at 774.

The approach of the *Tobey* court in interpreting the Privacy Act gives greater weight to public interest arguments and requires the judiciary to engage in the kind of intellectual heavy lifting necessary to resolve conflicts between access and privacy. While the Privacy Act is not a perfect parallel to FOIA, it is in many ways a superior model, for the Privacy Act was designed not to open governmental records, but to close them. That judges interpreting the Privacy Act recognize that information "about" individuals sometimes is first and foremost information "about" government is a subtlety demanded by these types of cases. Compared to the categorical approach of post-*Reporters Committee* FOIA cases, *Tobey* offers the promise of a more thoughtful framework for what has become a truly vexing issue.

The court's analysis in *Tobey*, though produced in the context of the Privacy Act, offers a useful method for resolving similar privacy-related questions in the FOIA context.